\documentstyle[12pt]{article}
\begin{document}
\baselineskip=24pt

{\Large Theory of Superconducting $T_{c}$ of doped fullerenes}

\vspace {1cm}

{\bf A.S. Alexandrov$^{*}$ and V.V. Kabanov$^{**}$ }

\vspace {1cm}

{\it $^{*}$Loughborough University of Technology, Loughborough LE11 3TU, U.K.

$ \it ^{**}$ IRC in Superconductivity, University of Cambridge, Cambridge CB3 
OHE, U.K. and Frank Laboratory of Neutron Physics, JINR, Dubna,Russia}

\vspace {1cm}

\begin{abstract}
We develop the nonadiabatic polaron theory of superconductivity of 
$M_{x}C_{60}$  taking into account the polaron band narrowing and 
realistic  electron-phonon and Coulomb interactions.
We argue  that the crossover from the BCS weak-coupling superconductivity 
to the strong-coupling polaronic and bipolaronic superconductivity occurs 
at the BCS coupling constant $\lambda\sim 1$ independent of the 
adiabatic ratio, and there is nothing ``beyond'' Migdal's theorem except 
small polarons for any realistic electron-phonon interaction.
 By the use of the polaronic-type  
function and the ``exact'' diagonalization  in the
truncated Hilbert space of vibrons (``phonons'') 
we calculate the ground state energy and the electron spectral density 
of the $C_{60}^{-}$ molecule. This allows us to describe 
the photoemission spectrum of $C_{60}^{-}$ in a wide energy 
region and determine the  electron-phonon interaction. 
The strongest coupling is found with the high-frequency  pinch $A_{g2}$ mode 
and with the Frenkel exciton.  We clarify the crucial role of 
high-frequency bosonic excitations in doped fullerenes which reduce the bare 
bandwidth and 
the Coulomb repulsion allowing  the intermediate and low-frequency  phonons  
to couple two small polarons in a Cooper pair.  The 
Eliashberg-type equations are solved  for low-frequency phonons.  
The  value of the 
superconducting 
$T_{c}$, its pressure dependence and the isotope effect
are found to be in a remarkable agreement with the available experimental data. 

\end{abstract}

\vspace {1cm}

PACS numbers 79.60.Bm, 63.20.-e, 71.38.+i

\newpage

\section {Introduction}

 Any phonon mediated superconductivity is a nonadiabatic phenomenon. 
 Nonadiabaticity of electrons in metals measured by the ratio of the 
 characteristic phonon frequency to the Fermi energy is normally small 
 ($\leq 10^{-2}$), so is their superconducting $T_{c}$. 
 High temperature superconductivity of  doped fullerenes \cite{exp,ram} 
seems to be due to large  nonadiabaticity. The phonon frequencies are high, 
$\omega \leq 0.2eV$ while  
the bare Fermi energy is very low  $E_{F}\leq 0.2eV$ \cite{pic}. Tolmachev 
$logarithm$ in the definition of the Coulomb pseudopotential $\mu^*$ does 
not apply in this nonadiabatic case and the  electron-phonon coupling should 
be strong ($\lambda \geq 1$) to overcome the Coulomb repulsion. The strong 
electron-phonon interaction implies  the nonadiabatic 
polaron or bipolaron superconductivity as discussed in detail by 
Alexandrov and Mott\cite{ale}.  However, the final answer on the origin of high 
$T_{c}$ depends not only on 
the adiabatic ratio $\omega/E_{F}$ but also on  the coupling constants 
with different phonons . If a relatively 
weak coupling ($\lambda\leq 0.5$) with low-frequency phonons dominates and  
the Coulomb repulsion is not very large, then
the Migdal-Eliashberg theory is  applied with the BCS ground  state. 
On the other hand, if the coupling is strong and (or) high-frequency  
phonons are involved, the nonadiabatic polaron theory  
is more appropriate. 

In the past years several different  calculations of the
electron-phonon coupling constants have been reported for fullerenes\cite 
{pic1,antr,var,shlu,faul}. Some of them yield the 
strongest coupling with the high frequency $H_{g}$ modes with a moderate 
 electron-phonon coupling, $\lambda \leq 0.5$.
On the other hand Picket $et$ $al$\cite {pic1} predicted  the strongest coupling 
 with the high frequency $A_g(2)$ 
mode and $\lambda \sim 3$. The difference 
in calculated coupling constants  is 
quite remarkable, and may result in a qualitatively different understanding of 
the nature of superconductivity of doped fullerenes.
Therefore, the experimental determination of  $\lambda$ and more 
extensive theoretical work are required.  
 
 Recent photoemission spectroscopy (PES) of the molecule $C^{-}_{60}$ \cite{gun}  
allows us to estimate the relative contribution of different  
phonons.  
While the variations analysis by Gunnarsson $et$ $al$  
in the truncated space up to five phonons showed 
the strongest coupling with a $low$-frequency $H_{g}$ mode, 
 we found by applying the polaron theory of PES  
 that the coupling with the high-frequency $A_{g2}$ mode dominates 
 \cite{alekab}.  The 
coupling constant with this mode appears at least by factor $\sim 1.5$ larger than with 
any other mode in a qualitative agreement with the tight-binding analysis by Picket 
$et$ $al$\cite{pic1}. This demonstrate that 
an estimate of the  electron-phonon coupling constants 
from PES, the Raman and neutron scattering  using the canonical 
theory of metals may be inconclusive,
since doped fullerenes are not conventional metals to which 
Migdal's ``theorem'' can  be applied. 

There is some confusion in the literature concerning the violation of 
Migdal's theorem. As was realized long time ago \cite{all} the Migdal 
\cite{mig} theory of electron-phonon effects in the normal state and the 
Eliashberg theory \cite{eli}, which generalizes BCS theory to incorporate 
Migdal theory in the normal limit, provide a rigorous basis for 
understanding $T_{c}$ in the weak and  intermediate coupling regime. 
By the use of the $1/\lambda$ expansion technique\cite{Ale} one can readily show 
that small polarons appear  at $\lambda\sim1$ and thus Migdal's 
theorem fails. The breakdown of the Migdal-Eliashberg theory at 
$\lambda\sim1$ has nothing to do with the nonadibatic corrections due to 
the so-called ``crossing'' diagrams (vertex corrections). Small polarons 
are formed at $\lambda \geq 1$ both in the initially adiabatic $\omega \ll E_{F}$ or 
nonadiabatic $\omega \geq E_{F}$ system irrespectively of the ratio 
$\omega/E_{F}$\cite{hol}. Their formation is the result of the broken 
translation symmetry as discussed in ref.\cite{alemaz,kab}. 
Of course, in  special cases
Migdal's theorem will not hold  even for small 
$\lambda <1$: first, if either characteristic phonon has small momentum 
$q \ll q_{D}$ ($q_{D}$ is the Debye momentum), 
and second, if the Fermi surface has a nested topology. In these textbook 
examples \cite{all,gin} the ``crossing'' diagrams are no longer small. 
Several  authors \cite{gra,shr,tak,gri}  suggested a theoretical treatment of the  vertex 
corrections with quite  opposite conclusions about their role for 
$T_{c}$. While many authors generated the opinion that vertex corrections 
should, in general, be negative, Grimaldi $et$ $al$ \cite{gri} obtain a 
strong enhancement of $T_{c}$ from non-adiabatic terms. 

In this paper we develop the theory of the superconducting $T_{c}$ which 
 takes into account the nonadiabatic effects, strong coupling with 
some phonons, and the 
realistic Coulomb interaction. First, we discuss in more details
the origin of the 
breakdown of Migdal's theorem due to the broken translation symmetry. We 
argue that while a strong enhancement of $T_{c}$ due to vertex corrections 
\cite{gri} is an artifact of an unrealistic  electron-phonon 
interaction, the polaronic band-narrowing provides such  enhancement quite 
naturally.  It means that the  multiphonon 
dressing is  important for high-$T_{c}$ superconductivity as was 
predicted by one of us \cite{ale2}. Then we analyze  the $PES$ 
data  \cite{gun} for $C_{60}^{-}$ using the   polaronic displacement transformation 
for $A_{g2}$ mode and the truncated Hilbert space for other phonons. Our fit 
to the experimental PES   is just as good as the variations 
approach \cite{gun} for the low binding energies and  better than that for the 
high-energy region. We obtain  the strongest coupling with the high-lying $A_{g}(2)$ 
pinch mode and with the Frenkel-type excitons.
 As a result, we determine the $bare$ coupling constants to formulate and 
 solve the 
 Eliashberg-type equations fully taking into account multiphonon dressing 
 and nonadiabatic motion of 
 carriers. These provide us with the value of $T_{c}$, its pressure 
 dependence and with the isotope effect 
 which are in excellent agreement 
 with experiment. In conclusion we analyze a possibility for the bipolaron 
 formation and $2e$ Bose-liquid ground state of doped fullerenes.

 \section{Broken translational symmetry and breakdown of \\ 
   Migdal's theorem.}

\noindent In doped semiconductors the carriers become small polarons or 
bipolarons at the intermediate value of the coupling constant 
$\lambda\geq 1$ or $\lambda\geq 0.5$, respectively \cite{alemot2}. However, within 
the adiabatic Migdal description of  electrons and phonons 
coupled by the linear electron-phonon interaction there is no instability 
 at any value of $\lambda$ if the bare 
 ionic plasmon mode is replaced by  the acoustic phonon mode\cite{all}.
The corrections to the normal state spectrum due to 
the coupling are adiabatically small
 ($\sim \omega/E_{F}$). In particular, the critical temperature of the 
 BCS superconductor is adiabatically small, $T_{c}/E_{F}<\omega/E_{F}\ll 1$ 
 for all relevant values of $\lambda$. Therefore, the self-consistent 
 Migdal-Eliashberg approach does not allow for the possibility to 
 study  the small-(bi)polaron formation in the intermediate and strong-coupling regime.
  This drawback is due to the basic assumption of the 
 canonical theory that the Green's functions (GFs) are translationally invariant, thus $G({\bf 
 r},{\bf r'},t)=G({\bf r}-{\bf r'},t)$. This assumption excludes the 
 possibility of the local violation of the translational symmetry due to 
 the lattice deformation followed by the self-trapping. To enable 
 the electron to relax into the lowest polaron state, one can introduce 
an infinitesimal translationally noninvariant 
 potential, which should be set equal zero only in the final solution for 
 GFs \cite{alemaz}. As in the case of the off-diagonal 
 superconducting order parameter, a small potential violating a 
 translational symmetry drives the system into a new ground state at 
 sufficiently large $\lambda$. Setting it equal to zero in the solution 
 to the equation of motion restore the translational symmetry but in a 
 new polaronic band rather than in the electron one, which turns out to 
 be an excited state.

In the Holstein model, in which electrons interact with the 
local (molecular) phonons one can notice 
the polaronic instability of the Fermi liquid at  $\lambda\sim 1$ 
 already with 
  Migdal's diagrams for the electron
self-energy. The interaction is described by the Hamiltonian
\begin{equation}
H_{e-ph}={1\over{\sqrt{2N}}}\sum_{\bf q,k}\gamma({\bf q}) \omega({\bf q}) 
c^{\dagger}_{\bf k+q}c_{\bf k} d_{\bf q} +H.c.,
\end{equation}
where the coupling constant $\gamma({\bf q})$ and the phonon frequency 
$\omega({\bf q})$ are momentum independent.
The electron self-energy $\Sigma$ in the Migdal approximation contains 
two contributions, $\Sigma_{M}$, Fig.1a and $\Sigma_{\mu}$, Fig.1b. 
  $\Sigma_{M}\simeq 
\lambda\omega$ and therefore
remains adiabatically  small compared with the  bandwidth $2D\simeq N(0)^{-1}$ in 
the relevant region of the coupling ($\lambda<D/\omega$), which 
guarantees the self-consistency of the approach. On the other hand for the
 molecular phonons
$\Sigma_{\mu}\simeq D\lambda n$ is not small and it turns out to be comparable or larger than the 
Fermi energy already at $\lambda\sim 1$ for any filling of the band 
 ($n$ is the electron density per cell).
As a rule  this diagram, which is momentum 
and frequency independent, is included in the definition of the 
chemical potential $E_{F}$. While this is justified  for a 
weak-coupling regime,   $\Sigma_{\mu}$ 
leads to an instability for a strong coupling. To show this let us consider a one-dimensional 
chain in the tight-binding approximation with the nearest-neighbor 
hopping integral $D/2$. The renormalized chemical potential is given by
\begin{equation}
E_{F}=D sin\left({\pi(n-1)\over{2}}\right)-2D\lambda n,
\end{equation}
with $\lambda=\gamma^{2}\omega/2D$. The system is stable if the 
derivative  
$dE_{F}/dn$ is positive, which yields the following region for the 
stability of the Migdal solution:
\begin{equation}
\lambda<{\pi\over{2}} cos\left({\pi(n-1)\over{2}}\right).
\end{equation}
For  two and three-dimensional lattices the numerical coefficient is 
different, but
the critical value of $\lambda$ remains  of the order of unity. 

In doped 
semiconductors and metals  phonons are  screened and the diagram 
$\Sigma_{\mu}$ containing only zero momentum phonons   vanishes.
Therefore, one can erroneously conclude that there is nothing to worry 
about as far as the applicability of the Migdal approach. 
However, it is sufficient to violate the translation symmetry and then 
restore it  to observe 
the instability of the bare band.  
The polaronic instability is essentially an 
adiabatic effect \cite{hol,kab}. Therefore, to see how the same self-energy 
  $\Sigma_{\mu}$ leads to  the polaronic collapse independent of 
the type of phonons we consider the extreme adiabatic limit of the 
classical deformation field $\phi({\bf r})$ coupled with the electron 
field $\psi({\bf r})$ as
\begin{equation}
H=H_{e}+\int d{\bf r}d{\bf r'}\left[g({\bf r}-{\bf r'})\phi({\bf r'}) 
\{\psi^{\dagger}({\bf r})\psi({\bf r})-n^{0}({\bf r})\}+h.c.\right]+s^{2}|\nabla 
\phi({\bf r})|^{2}.
\end{equation}
Here $H_{e}$ is the electron kinetic energy, $s$ is the sound velocity, $g({\bf r})$ 
is the coupling constant 
with the  Fourier component $g_{\bf q}=\omega_{\bf q}^{3/2}\gamma({\bf 
q})/\sqrt{2N}$,  $n^{0}({\bf r})$ is the periodic 
density of carriers respecting the translational symmetry, and $\omega_{\bf q}=sq$. 
Minimizing Eq.(4) with respect to the classical field $\phi^{*}({\bf 
r})$ we obtain
\begin{equation}
s^{2}\nabla^{2}\phi({\bf r})=\int d{\bf r'}g^{*}({\bf r}-{\bf 
r'})\{n({\bf r'})-n^{0}({\bf r'})\}.
\end{equation}
The solution is
\begin{equation}
\phi({\bf r})=-\sum_{\bf q}{g^{*}_{\bf q}\over{s^{2}q^{2}}}e^{i{\bf q}\cdot {\bf r}}
\{n_{\bf 
q}-n^{0}_{\bf q}\},
\end{equation}
where $n_{\bf q}$ is the Fourier component of the electron density 
$n({\bf r})\equiv \langle \psi^{\dagger}({\bf r})\psi({\bf r})\rangle$. 
Substituting Eq.(6) into the Hamiltonian Eq.(4) we find that the adiabatic 
lattice deformation leads to the lowering of the electron energy by the 
value
\begin{equation}
\delta \mu({\bf r})=-2\sum_{\bf q}{|g_{\bf q}|^{2}\over{\omega^{2}_{\bf 
q}}}e^{i{\bf q}\cdot {\bf r}}
\{n_{\bf 
q}-n^{0}_{\bf q}\}.
\end{equation}
If the electron density is periodic, $n_{\bf q}=n^{0}_{\bf q}$ 
the shift of the energy is zero. Otherwise, it is not.
 For example, one can consider a random statistically uncorrelated 
distribution with the $ansamble$ average $\langle{n_{\bf q} n_{\bf 
q'}}\rangle=Nn^{2}\delta_{{\bf q},{-\bf q'}}$. In that example 
 the chemical potential is  shifted by the value  
\begin{equation}
\langle|{\delta \mu}|\rangle=n\left[{1\over{N}}\sum_{\bf q}|\gamma({\bf 
q})|^{4}\omega_{\bf q}^{2}\right]^{1/2}\simeq 2n\lambda D,
\end{equation}
which is practically the same as in Eq.(2). However, now 
\begin{equation}
\lambda D \equiv E_{p}={1\over{2N}}\sum_{\bf q}|\gamma({\bf q})|^{2}\omega_{\bf q}
\end{equation}
 depends on the phonon spectrum, 
integrated over $all$ Brillouin zone rather than on  zero momentum 
phonons. The broken translational symmetry lowers the energy by the 
value $2E_{p}\simeq 2\lambda D$ per particle. The corresponding increase of the 
deformation energy is $E_{p}$. Therefore, the system prefers to relax 
into the self-trapped state if $E_{p}>D$.
The nonadiabatic vertex corrections neglected within the Migdal 
approach are also completely irrelevant for the small polaron formation 
and, therefore for the breakdown of this approach.
The extension of the 
Migdal approximation to the strong-coupling limit  is 
unacceptable because of the broken translation symmetry.

\section {Exact nonadiabatic solution  
 and polaron band \\ restoring  translational symmetry}

\noindent 
In the small polaron regime, $\lambda\geq 1$ 
the kinetic energy remains smaller than the 
 interaction energy and a selfconsistent 
 treatment of a many polaron system  is 
 possible with a 
 $'1/\lambda'$ expansion technique  in the nonadiabatic 
 regime\cite{Ale}. This 
 possibility results from the fact, known for a long time, that there is 
 an exact solution for  
a single electron without hopping term. Following  Lang and Firsov \cite{lan} one can apply
the canonical transformation  to diagonalise the Hamiltonian in the site 
representation
\begin{equation}
H=\sum_{<ij>}T({\bf m-n})c^{\dagger}_{i}c_{j}+\sum_{\bf q}\omega({\bf 
q})(d^{\dagger}_{\bf q}d_{\bf q}+1/2)+\sum_{{\bf q},i}\omega({\bf 
q})n_{i}[u_{i}({\bf q})d_{\bf q}+H.c.]+H_{c},
\end{equation}
where $T({\bf m})$ is the bare hopping integral, $i=({\bf m},s)$, $j= 
({\bf n},s)$ ($s$ stands for spin),
$u_{i}({\bf q})=\gamma({\bf q})exp(i{\bf q}\cdot {\bf m})/\sqrt{2N}$, and $H_{c}$ 
is the $direct$ Coulomb repulsion. 
 The diagonalisation 
is exact if $T({\bf m})=0$ (or $\lambda=\infty$):
\begin{equation}
\tilde {H}=e^{S} H e^{-S},
\end{equation}
where
\begin{equation}
S=\sum_{{\bf q}, 
 i}\hat{n}_{i}\left(u_{i}({\bf q})d_{\bf q} -h.c.\right).
 \end{equation}
  The electron operator transforms as
 \begin{equation}
 \tilde{c}_{i}=c_{i} exp\left(\sum_{\bf q}u_{i}({\bf q})d_{\bf 
 q}-h.c.\right)
 \end{equation}
 and  the phonon one as
 \begin{equation}
 \tilde{d}_{\bf q}= d_{\bf q}-\sum_{i}\hat{n}_{i}u^{*}_{i}({\bf 
 q})
 \end{equation}
 From Eq.(14) it follows that the Lang-Firsov canonical transformation 
 is the displacement transformation 
for the multi-polaron system shifting  ions to the new equilibrium positions. In a more general sense it 
 changes the boson vacuum. As a result:
 \begin{equation}
 \tilde{H}=\sum_{i,j}(\hat{\sigma}_{ij}-\mu\delta_{i,j})c^{\dagger}_{i}c_{j}
 -E_{p}\sum_{i}\hat{n}_{i}+
 {1\over{2}}\sum_{i,j}v_{ij}c^{\dagger}_{i}c^{\dagger}_{j}c_{j}c_{i}+\sum_{\bf 
 q}\omega_{\bf q}(d_{\bf q}^{\dagger}d_{\bf q}+1/2)
 \end{equation}
where
\begin{equation}
\hat{\sigma}_{ij}=T({\bf m-n})\delta_{s.s'}exp\left(\sum_{\bf q} 
 [u_{i}({\bf q})-u_{j}({\bf q})]d_{\bf q} -h.c.\right)
 \end{equation}
 is the new hopping integral depending on the phonon variables,
  \begin{equation}
 v_{ij}= V_{c}({\bf m-n})-2\sum_{\bf q}\omega_{\bf q} \left(u_{i}({\bf 
 q})u^{*}_{j}({\bf q})\right)
 \end{equation}
 is the polaron-polaron interaction comprising the  direct Coulomb 
 repulsion $V_{c}$ and 
 the attraction via a $nonretarded$ lattice deformation (second term in 
 Eq.(17)).
 In the extreme strong-coupling limit $\lambda\rightarrow\infty$ one can 
 neglect the  hopping term of the transformed Hamiltonian. The rest 
 has analytically determined eigenstates and eigenvalues. The eigenstates 
 $|\tilde{N}\rangle=|n_{i}, n_{\bf q}\rangle$ are 
 classified with the polaron  $n_{{\bf 
 m},s}$
  and phonon $n_{\bf q}$ occupation numbers  and the energy levels are:
  \begin{equation}
  E=(T(0)-E_{p}-\mu)\sum_{i}n_{i}+{1\over{2}}\sum_{i,j}v_{ij}n_{i}n_{j}+\sum_{\bf 
  q}\omega_{\bf q}(n_{\bf q}+1/2)
  \end{equation}
  where $n_{i}=0,1$ and $n_{\bf q}=0,1,2,3,....\infty$. The interaction 
  term does not include the on-site interaction $i=j$  for parallel 
  spins because of the Pauli principle. The polaronic (Franck-Condon) 
  level shift is expressed in a general form independent on the type of 
  phonons in Eq.(9).
 
  Thus we conclude that the Hamiltonian Eq.(10)
  in zero order of the hopping 
  describes  localized polarons and independent phonons 
  which are vibrations of ions relative to new equilibrium positions 
  depending on the polaron occupation numbers. The phonon frequencies 
  remain unchanged in this limit\cite{alecap,kab}. The middle of the electronic band 
  $T(0)$ falls down by $E_{p}$ as a result of a potential well produced 
  by the lattice deformation due to the self-trapping. With the finite hopping term  polarons tunnel in a narrow band
because of the degeneracy of the zero order Hamiltonian relative 
the site position of a single polaron in a regular lattice. To show this
self-consistently one can apply  perturbation theory using $1/\lambda$ 
as a small parameter.  
Because of the degeneracy terms of the first order in 
$T({\bf m})$   should be included in a zero order 
Hamiltonian $H_{0}$:
\begin{equation}
H_{0}=\sum_{i,j}(\sigma({\bf m-n})-\mu\delta_{i,j})c^{\dagger}_{i}c_{j}+\sum_{\bf 
 q}\omega_{\bf q}(d_{\bf q}^{\dagger}d_{\bf q}+1/2)
 \end{equation}
where
\begin{equation}
\sigma({\bf m-n})=\langle\langle \hat{\sigma}_{ij}\rangle\rangle=
T({\bf m-n})\delta_{s,s'}exp[-g^{2}({\bf m-n})]
\end{equation}
is the hopping integral averaged with the phonon equilibrium distribution.
It is  calculated 
by the use of the relation
\begin{equation}
e^{A+B}=e^{A}e^{B}e^{-[AB]/2},
\end{equation}
 For  zero temperature one obtains 
\begin{eqnarray}
\sigma({\bf m-n})&=&T({\bf m-n})e^{-g^{2}({\bf m}-{\bf n})}\cr
&\times&\langle 0|exp\left[{-\sum_{\bf q}[u_{i}^{*}({\bf q})-u_{j}^{*}({\bf q})]
d^{\dagger}_{\bf 
q}}\right] exp \left[{\sum_{\bf q}[u_{i}({\bf q})-u_{j}({\bf q})]d_{\bf 
q}}\right]|0\rangle,
\end{eqnarray}
where
\begin{equation}
 g^{2}({\bf m}-{\bf n})
 ={1\over{2}}\sum_{\bf q}\left(|u_{i}({\bf q})|^{2}+|u_{j}({\bf 
 q})|^{2}-2u_{i}^{*}({\bf q})u_{j}({\bf q})\right).
 \end{equation}
The bracket  is equalto unity for $T=0$. The straightforward 
generalization for finite temperatures \cite{hol,lan} yields
\begin{equation}
g^{2}({\bf m})={1\over{2N}}\sum_{\bf q}|\gamma({\bf q})|^{2}
coth\left({\omega_{\bf q}\over{2T}}\right)[1-cos({\bf q\cdot m})].
\end{equation}
  Earlier we have  shown \cite{alekab2} by the numerical diagonalization 
  of a finite site Holstein model that the expression for the polaron 
  bandwidth, Eq.(20) is valid for the nonadiabatic and intermediate regime 
  $\omega \geq D$ for $all$ coupling strength. In 
  the adiabatic case important corrections 
  appear. However, the main exponential term remains almost the same 
  in the 
  strong coupling regime, $\lambda >>1$.
  As a result of this brief recourse to the multi-polaron theory, one can 
  conclude that the major effect of high-frequency phonons is the 
  reduction of the bandwidth and of the Coulomb repulsion. While optical 
  phonons cannot overscreen the repulsion they reduce its value 
  significantly. Then the acoustical and and low-frequency molecular modes provide a net 
  attraction. 
  
  It was argued by one of us \cite{ale2} that the polaron band 
  narrowing  can provide a high value of $T_{c}$ because of the 
  enhancement of the 
  density of states. This enhancement does not depend 
  on the particular choice of the electron-phonon interaction contrary to
  the enhancement of $T_{c}$ due to the vertex corrections 
  \cite{gri}. The latter can be obtained only if the electron-phonon scattering is 
  dominated by small momentum transfer $q \ll q_{D}$. A simple estimate of 
  the screening radius in doped fullerenes and cuprates yields a value of 
  the order or even less than  the lattice constant. Therefore, a long-range 
  electron-phonon interaction is ruled out by screening, which makes the 
  positive effect of the vertex corrections quite unrealistic \cite{dol}.

\section{PES and bare electron-phonon interaction}
 
 Because of the diversity in the theoretical results for  
 electron-phonon interaction the coupling strength can be  unambiguously determined only 
 from the experiment. 
Allowing  a small shift of the phonon frequencies due to screening the 
 coupling with the 
 intraball modes are expected to be the same in doped fullerenes as in the 
 $C_{60}^{-n}$ molecule. 
 The recent
photoemission spectroscopy of a molecule $C^{-}_{60}$ \cite{gun}
 allows us to estimate the relative contribution of different 
phonon modes and other bosonic excitations 
to the interaction. 

In this section we analyze the $PES$ data  \cite{gun} using
the  exact polaronic
diagonalization with respect to  the $A_{g}(2)$ mode  
and introducing 
the $polaron-exciton$ coupling.    We obtain  
  the strongest 
   coupling with the high-lying $A_{g}(2)$ pinch mode and with a 
   Frenkel-type exciton. As a result we provide a strong evidence for the 
   nonadiabatic  coupling with high-energy bosonic excitations in 
   $M_{x}C_{60}$.
       
    The Hamiltonian at hand, describing three degenerate $t_{1u}$ states coupled with 
    phonons, is diagonalized with respect to the $A_{g2}$ 
    coupling using the 
    canonical Lang-Firsov displacement transformation as described above.
     The result is
   \begin{equation}
     \tilde{H} =e^{S}He^{-S}=
     -E_{p}^{A_{g2}}\sum_{m=1}^{3}\psi_{m}^{\dagger}\psi_{m}+\sum_{\nu=1}^{8}
     g^{\nu}\omega_{\nu}\sum_{n,m=1}^{3}\psi_{n}^{\dagger}M_{nm}^{\nu}\psi_{m}
     +\sum_{\nu}^{8}\sum_{\mu=1}^{5}\omega_{\nu}n_{\nu,\mu},
     \end{equation}
     where $E^{A_{g2}}_{p}=g^{2}\omega_{A_{g2}}$ 
     is the polaron shift due to the $A_{g2}$ mode,
     $3\times3$ dimensionless  matrix $\hat{M}$  is taken from ref.\cite{lan}
     \begin{displaymath}
     \hat{M}=\left( \matrix {\sqrt{3}Q_{4}+Q_{5} & \sqrt{3} Q_{1} & \sqrt{3} 
     Q_{2}\cr \sqrt{3}Q_{1} & -\sqrt{3} Q_{4}+Q_{5} & \sqrt{3} 
     Q_{3}\cr \sqrt{3}Q_{2} & \sqrt{3} Q_{3} & -2
     Q_{5}}\right),
     \end{displaymath}
     and $n_{\nu,\mu}$ are the phonon occupation numbers of eight 
     five-degenerate $H_g$ modes 
     with the phonon operators $Q^{\nu}_{\mu}=d^{\dagger}_{\nu,\mu}+d_{\nu,\mu}$. 
    The interaction with $H_{g}$ modes is responsible for the dynamic 
    Jahn-Teller effect in $C_{60}$. According to calculations \cite{gun2}
    singly ionised $C_{60}^{-}$ is in the intermediate coupling regime, 
    while the doubly and triple ionised molecules are in the strong 
    coupling limit with respect to the coupling with $H_{g}$ modes. 
    Therefore a reasonable estimate of the ground state energy is 
    obtained by taking into account only diagonal part of $\hat{M}$. 
    Nevertheless, to avoid any ambiguity we calculated 
    the spectral function $I_{pol}(\omega)$ of the 
     Hamiltonian, Eq.(25) by the exact numerical diagonalisation in truncated Hilbert space  
   (up to $4$ phonons) for the $H_g$ modes as described in 
   ref.\cite{alekab2}. A self-trapped 
     exciton in neutral $C_{60}$ was observed and discussed by several 
     authors \cite{mat,wan,jan}. Because of the polaron-exciton 
      coupling we add the same 
     spectral function to the total spectral density shifted by the 
     exciton energy $\omega_{ex}$, and multiplied by the polaron-exciton 
     coupling constant $g^{2}_{ex}$ as
     \begin{equation}
     I(\omega)=I_{pol}(\omega)+g^{2}_{ex} 
     I_{pol}(\omega+\omega_{ex}).
     \end{equation}
      This is an exact procedure if the 
    interaction with excitons is linear as with phonons. Then we  integrate $I(\omega)$ 
     with the Gaussian instrumental resolution function of width 
     $\sim 41 meV$ \cite{gun} taking into account the damping $\gamma_{ex}$ of the 
     exciton in the second (excitonic) contribution with the substitution
     $\omega_{ex}\rightarrow \omega_{ex}+i\gamma_{ex}$.
    We thus can fit 
     the $PES$ in a wide energy region  as shown in Fig.2 with  
     $g^{\nu}$  being the  fitting parameters (inset).  The 
     polaron-exciton coupling constant is found to be $g^{2}_{ex}=0.5$, the 
     exciton energy $\omega_{ex}\simeq 0.5 eV$ in agreement with the 
     luminescent data \cite{mat}. The inverse exciton lifetime is 
     estimated to be $\gamma_{ex}\simeq 580 cm^{-1}$.  The coupling to 
     the $A_{g}(2)$ mode turns out most important  in 
     agreement with the  tight-binding calculations \cite{pic} and with the
     doping 
dependence of the phonon frequencies and the Raman linewidths. The $A_g(2)$ mode 
shows a clear shift with doping from 1467$cm^{-1}$ in undoped $C_{60}$ to 
1446$cm^{-1}$ in $K_3C_{60}$ and to 1431$cm^{-1}$ in $K_6C_{60}$. The 
linewidth increases by $\sim 3cm^{-1}$ in the metallic state ($K_{3}C_{60}$) 
and  comes back to the value of $C_{60}$ with further doping in an 
insulator $K_{6}C_{60}$. At the same time 
high frequency $H_g(7,8)$ modes also show large broadening in the metallic 
state. However,  $Na_{x}C_{60}$, which does not exhibit metallic state with 
doping, shows the same strong line broadening and the bleaching of the 
$H_g(7,8)$ modes \cite {kuz}. Consequently, the broadening of phonon lines 
does not provide information about the electron-phonon coupling. Moreover, as 
has been pointed by Gelfand \cite {gel} $A_g(2)$ mode cannot decay into 
electron-hole pair, no mater how strong the electron-phonon coupling is. 
This decay is prohibited by the conservation of the energy because the 
estimated half-bandwidth is below the $A_{g}(2)$ frequency. We believe 
that an increase of the 
linewidth of $H_g$ modes in the metallic phase is associated with 
the  crystal field splitting of the five-fold degenerate mode. If this splitting is not 
very large it may result in an increase of the Raman linewidth. 
Therefore, the large broadening of $H_{g}$ modes does not contradict to 
our conclusion that their coupling with the electron is weak. 

In general, an estimate of the  electron-phonon coupling constants 
from the Raman  and  neutron scattering  using the canonical 
theory of metals may be inconclusive
since doped fullerides are not conventional metals. In particular, the 
Migdal ``theorem'' can not be applied.

\section {Polaron theory of $T_{c}$}

 It follows from the  PES analyses that the frequencies of essential bosonic excitations 
   (phonons and excitons) strongly coupled with electrons ($ g \sim 0.6 $)
     are above or of the same  order as the   electron half-bandwidth in  doped 
     fullerenes.
     This fact as well as the observation of the phonon and exciton-sided 
     bands in PES by itself favor
      the nonadiabatic small polaron theory \cite{ale} rather than
     the adiabatic Migdal-Eliashberg approach  to 
     $M_{x}C_{60}$. On the other hand the total contribution of intermediate and 
      low-frequency modes is not negligible, while their 
      individual coupling is relatively weak ($g\leq 0.3$). 
      This allows us to treat these modes within the Migdal-Eliashberg 
      approximation by the use of the Eliashberg function
 \begin{equation}
  \alpha^{2}F(\Omega)= {1\over{2}}\sum_{\nu=1}^{8}\lambda_{\nu} \omega_{\nu} 
  \delta(\Omega-\omega_{\nu})
 \end{equation}
 where $\nu=1,...8$ is the sum over 8 $H_{g}$ modes with
  \begin{equation}
  \lambda_{\nu} = 
 {20\over{3}}\cdot g_{\nu}^{2}\omega_{\nu} N(0),
 \end{equation}
 and $g_{\nu}, \omega_{\nu}$ from the inset in Fig.2. The additional coefficient $10$ 
 in the  partial  $\lambda_{\nu}$ appears due to the five-fold degeneracy and 
 the nondiagonal terms in 
 the matrix $\hat{M}$ as explained in ref. \cite{shlu}.
 At the same time the strong nonadiabatic interaction with the pinch mode 
 is fully taken into account by the renormalization 
 of the halfbandwidth $D$ as
 \begin{equation}
 W= D e^{-g^{2}}
 \end{equation}
 and by the reduction of the bare Coulomb repulsion as
 \begin{equation}
 \tilde{V}_{c}= 
 V_{c}-2g^{2}\omega_{A_{g2}}.
 \end{equation}
The excitonic contribution is taken into account via the 
high-frequency dielectric constant $\epsilon$. Because of the  covalent 
nature of  $C_{60}$ molecules we do not expect any significant 
dispersion of $\epsilon$, which is estimated as $\epsilon\simeq 4.4$ \cite 
{heb}. In a solid the exciton frequency $\omega_{ex}$ appears to be rather 
close to the fundamental gap $\sim 2 eV$. Therefore, we assume that the 
excitonic effect on the bandwidth is included in the LDA density of 
states (DOS) $N(0) = 1/2D$. In the following we take 
into account the polaron band narrowing due to the coupling with the 
pinch mode only. A possibility remains that we  overestimate this 
coupling at the expense of the coupling with the two high-frequency $H_{g}$ 
modes as discussed by Gunnarsson $et$ $al$ \cite{gun}. If so, one should 
also include the contribution of these two modes in the polaron band 
narrowing and in the reduction of the Coulomb repulsion by 
a simple redifinition of the high-frequency coupling constant $g$.  

  With $g^{2}\omega_{A_{g2}}\simeq 0.06eV$ (inset Fig.2) 
 we obtain the lowering of the Coulomb repulsion 
by  about $0.12 eV$. By the use of the Coulomb law,  the dielectric 
constant $\epsilon \simeq 4.4$ and the lattice constant $a\simeq 14 \AA$
 we estimate $intracell$ bare 
repulsion as $V_{c}\simeq 0.25 eV$. This value is in agreement with the 
binding exciton 
energy in solid $C_{60}$ \cite{kuz}. As a result, the residual repulsion is 
estimated as $0.13 eV$. The polaron density of states 
determined as
\begin{equation}
N_{p}(0)=N(0) e^{g^{2}}
\end{equation}
appears to be about $N_{p}(0)\simeq 9 states/eV spin$ if the bare 
density $N(0)\simeq 6.6 states/eV spin$ according to  some LDA calculations 
for $K_{3}C_{60}$ \cite{pic}. Because different LDA calculations yield 
rather different values of the bare $N(0)$ from $6.6$ up to $12.5 
states/eV spin$ \cite{gel} the polaronic density can be in the 
range from $9 states/eV spin$ up to about $17 states /eV spin$. 
These values of the polaron density of states are compared with those  
 measured with NMR   ($17 states/eV spin$), 
 thermopower ($14 states/eV spin$), and  magnetic susceptibility \\ ($14 
states/eV spin$)\cite{ram}, and estimated from the heat capacity jump and 
ESR \\ ($\sim 10 states/eV spin$) for $K_{3}C_{60}$. We note that different from the 
ordinary metals  the polaron heat capacity and the 
Pauli susceptibility are renormalized by the same amount in the 
polaronic system\cite{ale}.
By taking $6.6 states/eV spin $ as the bare total DOS of  three degenerate 
bands we estimate the bare intraband Coulomb pseudopotential as 
\begin{equation}
\mu={1\over{3}}N(0)V_{c}=0.55,
\end{equation}
and the renormalized one as
\begin{equation}
\mu^{*}={1\over{3}} N_{p}(0)\tilde{V}_{c}=0.4.
\end{equation}
Now we are prepared to solve the Eliashberg equations for $T_{c}$:
\begin{eqnarray}
Z(\omega_n)&=&1+8T {N_{p}(0)\over{N(0)}}\sum_{\omega_{n^{'}} > 0}
\omega_{n^{'}}tan^{-1}\left(\frac 
{W}{2\omega_{n^{'}}Z(\omega_{n^{'}})}\right) \cr
&\times&
\int_{0}^{\infty}d\Omega 
\alpha^{2}F(\Omega) \frac {2\Omega}{\left((\omega_{n^{'}}-\omega_{n})^{2}+
\Omega^{2}\right) 
\left((\omega_{n^{'}}+\omega_{n})^{2}+\Omega^{2}\right)}, \nonumber
\end{eqnarray}

\begin{eqnarray}
\Delta(\omega_n)&=&{T\over{Z(\omega_n)}}\sum_{\omega_{n^{'}}}
{2\Delta(\omega_{n'})\over{\omega_n^{'}}}tan^{-1}\left(\frac 
{W}{2\omega_{n^{'}}Z(\omega_{n^{'}}}\right)\cr
&\times& \left[{N_{p}(0)\over{N(0)}}\int_{0}^{\infty}d\Omega 
\alpha^{2}F(\Omega) \frac {2\Omega}{(\omega_{n^{'}}-\omega_{n})^{2}+
\Omega^{2}}-\mu^{*}\right]
\end{eqnarray}

First we show the normal DOS $\tilde N(E)$ in Fig.3 calculated with the analytical 
continuation of the normal state self-energy ($\sim Z-1$) to real 
frequencies. $\tilde N(E)$ takes into account the finite 
bandwidth,the band narrowing effect  and  
scattering by  low-frequency phonons. these results in 
drastic changes of the renormalized DOS compared with the initial DOS. 
 The latter 
is taken energy independent for simplicity. 

The result for $T_{c}$ is 
\begin{equation}
T_{c}= 20 K ,
\end{equation}
 for $K_{3}C_{60}$, and
 \begin{equation}
 T_{c}=35 K ,
 \end{equation}
 for $Rb_{3}C_{60}$. These values are obtained with the  Eliashberg 
 function, Eq.(27) and with the bare (LDA) DOS  $N(0)= 6.6 states/eV spin$ 
 and $N(0)= 7.5 states/eV spin$ for $K_{3}C_{60}$ and $Rb_{3}C_{60}$, respectively.
 
  Assuming 
 the inverse square root dependence of all intraball frequencies on the 
 carbon mass and the square root dependence of $g^{2}_{\nu}$ 
 we calculate the isotope effect. Normally, in the polaronic 
 superconductor the isotope exponent $\alpha= -d \ln {(T_{c})}/d \ln{(M)}$ is 
 negative because of the increase of the polaronic density of states with 
 the increase of the ion mass $M$  ( $g^{2}\sim \sqrt 
 {M}$ in Eq.(31)) \cite{ale4}. However,  the usual positive contribution to $\alpha$ 
 from low-frequency modes can change its sign, which is indeed the case 
 here. 
 By the use of the Eliashberg equations with the $isotope$ $mass$ 
 $dependent$ polaronic DOS, Eq.(31) we obtain
 \begin{equation}
 \alpha=0.27
 \end{equation}
 for $K_{3}C_{60}$,
 and
 \begin{equation}
 \alpha=0.31
 \end{equation}
 for $Rb_{3}C_{60}$. These values are rather close to those reported in 
 the literature \cite{pic}. 
 
 Finally, we calculate the dependence of $T_{c}$ on pressure $P$ assuming 
 that only  hopping integrals depend on pressure. In that case 
 the polaronic halfbandwidth $W$ has the same pressure dependence as the 
 bare one, which  according to ref. \cite{mart} is described by
 \begin{equation}
 W(P)=W(0) exp \left({5P\over{K}}\right),
 \end{equation}
with the elastic modulus $K\simeq 24.5 GPa$ \cite{zho}. Solving the 
Eliashberg equations with the pressure dependent bare and polaronic bandwidth 
we arrive with  the curve of Fig.4, which agrees well with the available 
experimental data.

\section {Conclusion}

A simple estimate of the characteristic energy scale of phonons and 
electrons in $M_{x}C_{60}$ clearly shows that the canonical 
Migdal-Eliashberg theory should be modified to include the polaronic 
nonadiabatic effects. As  discussed in the beginning of the paper 
the polaron collapse of the band  is the 
consequence of the broken translation symmetry and has nothing to do with 
the vertex corrections. The polaron band narrowing represents a universal 
origin of the high $T_{c}$ value. We argue that the high-temperature 
superconductivity of doped fullerenes is an example of the nonadibatic polaronic 
superconductivity discussed by one of us earlier \cite{ale2}. In that 
regime the polaron halfbandwidth is below the characteristic phonon 
frequency but still above the effective attraction between polarons. 
Therefore, 
the canonical BCS-Migdal-Eliashberg theory can be applied with the finite 
polaronic bandwidth 
and reduced Coulomb repulsion. The high frequency bosonic excitations 
play an important role reducing the bandwidth and the repulsion. 
They can be integrated out with the canonical displacement 
transformation, while the interaction with the low-frequency modes is 
still  described by the Eliashberg equations. As a result our  
theory gives the experimentally observed values of $T_{c}$, the isotope 
and pressure effects in $M_{x}C_{60}$ with the realistic coupling constants 
based on the PES data, the Coulomb law and common sense. 

Our calculated polaron halfbandwidth is about $W\simeq 150 meV$, or less, Fig.3. 
The total contribution of all vibration modes to the attraction is 
 $2E_{p}^{total}\simeq 330 meV$ (insert in Fig.2). As a 
result the  attractive interaction between two small polarons is 
estimated as $\Delta=2E_{p}-V_{c}\simeq 80 meV$. The condition $W\geq 
\Delta$,  necessary for the Fermi-liquid approach  seems to be  
satisfied. However, the system is very close to the bipolaronic instability 
$\Delta>W$ with the 2e Bose-liquid ground state. The final answer on the 
nature of the ground state remains with the experiment. The 
 hallmarks of small bipolarons are  those of small polarons, plus
 
 - superfluid phase transition similar to that  of $He^{4}$,
 
 - spin gap in the magnetic susceptibility, 
 that is $\chi_{s}\rightarrow 0$ at $T\rightarrow 0$
  if a singlet is the 
 ground state, and the absence of the Hebel-Slichter peak in the NMR,
 
 - electrodynamics of the charged Coulomb Bose gas with the divergent 
 $H_{c2}(T)$,
 
 - double elementary charge $2e$ in the normal state. 
 
 The first three features  are clearly observed in doped cuprates 
 \cite{ale}. The upward $H_{c2}(T)$ curvature near $T_{c}$ is measured  
 also in $M_{x}C_{60}$ as discussed in Ref.\cite{ale5,pic}.

We acknowledge helpful discussions with Sir Nevill Mott,  O. Dolgov, J.Ranninger,
E. Salje, and J. Samson,  
and the Royal Society financial support for one of us (V.V.K.).

\newpage

{\Large Figures} 

Fig. 1 Self-energy in Migdal's approximation.

Fig. 2 Polaron theory fit (full line) to the experimental $PES$ (dashed 
line). Frequencies $\omega =\omega_{\nu}$, 
 coupling constants $g = g^{\nu}$, and the contribution to the ground state energy $E 
 = E_p^{\nu}$ for different modes are shown in the inset.   
 For comparison we also show the coupling constants (g(Gun), inset) and 
 the calculated variations
 PES (dotted line) of ref.\cite {gun}.
 
Fig. 3 Polaron density of states in $K_{3}C_{60}$. The bare DOS is 
 shown with the dashed line.
 
Fig. 4 Pressure dependence of $T_{c}$ for $K_{3}C_{60}$  and 
 $Rb_{3}C_{60}$ \cite{sparn} compared with the theory.

\end{document}